\documentclass[aps,floatfix,showpacs,twocolumn,prl]{revtex4}
\usepackage[dvips]{graphicx}
\usepackage{amsfonts}
\usepackage{amssymb}\usepackage{times}
\usepackage{sub_JP}
\usepackage{MHequ}

\def\L{{\rm L}}
\def\R{{\rm R}}

\def\j{{\rm j}}

\def\G{{\rm G}}

\def\rho{\varrho}

\def\eref#1{(\ref{#1})}

\def\a{\alpha}
\def\b{\beta}
\def\b{\gamma}
\def\e{\varepsilon}
\def\v{\nu}
\def\r{\varrho}
\def\aG{\a_\G}
\def\bG{\b_\G}
\def\aLL{\a_\L^-}
\def\aLR{\a_\L^+}
\def\aRL{\a_\R^-}
\def\aRR{\a_\R^+}
\def\bLL{\b_\L^-}
\def\bLR{\b_\L^+}
\def\bRL{\b_\R^-}
\def\bRR{\b_\R^+}
\def\eLm{{\e_\L^-}}
\def\eRm{{\e_\R^-}}
\def\eLp{{\e_\L^+}}
\def\eRp{{\e_\R^+}}
\def\vLm{{\v_\L^-}}
\def\vRm{{\v_\R^-}}
\def\vLp{{\v_\L^+}}
\def\vRp{{\v_\R^+}}
\def\A{{\mathcal{A}}}
\def\B{{\mathcal{B}}}
\def\Am{{\mathcal{A}^-}}
\def\Ap{{\mathcal{A}^+}}
\def\Bm{{\mathcal{B}^-}}
\def\Bp{{\mathcal{B}^+}}
\def\AA{{\mathcal A}}

\def\e{{\varepsilon}}
\def\LR{{\stackrel{\mathrm L}{\scriptscriptstyle{\R}}}}
\def\ram{{\varrho_\alpha^-}}
\def\rap{{\varrho_\alpha^+}}
\begin{document}

\title{Thermal Rectification in Billiard-like Systems}
\author{Jean-Pierre Eckmann${}^{1,2}$}
\author{Carlos~Mej\'\i{a}-Monasterio${}^1$}
\affiliation{${}^1$D\'epartement de Physique Th\'eorique, Universit\'e de Gen\`eve}
\affiliation{${}^2$Section de Math\'ematiques, Universit\'e de Gen\`eve}
\date{\today}
\begin{abstract}
  We study  the thermal rectification phenomenon in  ``billiard'' systems with
  interacting particles.  This interaction  induces a local dynamical response
  of the  billiard to  an external thermodynamic  gradient. To  explain this
  dynamical  effect we study  the steady  state of  an asymmetric  billiard in
  terms  of the  particle and  energy reflection  coefficients. This
  allows us to
  obtain expressions for the region in parameter space where large thermal
  rectifications  are  expected.   Our  results  are  confirmed  by  extensive
  numerical simulations.
\end{abstract}

\maketitle

\noindent{\bf Introduction}. The  emergence  of  macroscopic   behavior  from  the  underlying  microscopic
dynamical laws is a major puzzle of theoretical physics.  However, despite its
long history, attempts to link  these macroscopic laws to
the  underlying  microscopic  dynamics  have  not  been  conclusive  thus  far
\cite{bonetto}.

With  the advent  of nanotechnology,  the study  of dynamics  of systems  at a
nanoscopic scale starts to play a  central role. In view of the broad spectrum
of  possible applications,  the investigation  on  how the  properties of  the
microscopic  dynamics can  be  used to  understand  and thus,  to control  the
behavior of nano-devices has attracted much attention \cite{nanoscopics}.

One such example  is the phenomenon of thermal  rectification (TR).  A thermal
rectifier is  a device in which the  magnitude of the heat  current depends on
the sign  of the imposed  temperature gradient.  In  the past there  have been
several  experimental  attempts  to design  such  a  device  (see {\em  e.g.},  
\cite{rectifier} and  references therein).   However, a possible
theoretical mechanism to explain the phenomenon from first 
principles has been proposed only recently \cite{terraneo}.

So far,  there is  no unique  framework to explain  the phenomenon  of thermal
rectification.   In  \cite{terraneo} and  the  later papers  \cite{li-2,li-3},
different models  of anharmonic oscillators  coupled to stochastic  heat baths
have  been studied.   Because  of  anharmonic terms,  the width  and
position of an ``effective'' phonon band depend on temperature.  It has been
shown  that by tuning  the anharmonicities  of two  or more  different coupled
chains  it is possible  to control  the effective  overlap between  the phonon
bands of  the two chains.   In particular, the  extent of this overlap  can be
made to depend on the sign of  the imposed thermal gradient, thus leading to a
TR.   A  further  step  to  devise  a thermal  diode  has  been  discussed  in
\cite{li-3} in terms of  the negative differential thermal resistance observed
in  some anharmonic  chains.  More recently,  in \cite{segal}  a  thermal rectifier
based  on  a  spin-boson  nanojunction  model has  been  discussed.

In  this  Letter we  study  the phenomenon  of  thermal  rectification in  low
dimensional  ``billiard'' Hamiltonian  systems. Roughly  speaking,  a billiard
consists of an  \emph{ensemble} of free particles colliding  elastically with a
fixed boundary.  Furthermore, to  study heat transport and thus rectification,
we  consider that  the billiard  is coupled  at two  different openings  of its
boundary with two different thermodynamic baths.

In a billiard, as described above,  TR is identically zero irrespective of the
geometry of  the boundary, as the  strengths of the fluxes  inside the billiard
are determined  only by its  coupling with the  external baths.  Here,  we show
that  if the particles are  allowed to interact (as  defined below), then
rectification of the heat flux is possible, and can be quite large.

The model we consider consists of a gas of non-interacting point particles of
mass $m$ that move freely inside a one-dimensional open channel made of $2$
identical two-dimensional cells (like chaotic billiards), connected through
openings of size $\delta$.  At the center of each cell, there is a fixed,
freely rotating disc of radius $R$ with which particles interact, exchanging
energy as follows: when a particle hits a disc, it exchanges its tangential
velocity $v_t$ with the angular velocity $\omega $ of the disc, while the
normal component $v_n$ is reflected \cite{mejia}:
\begin{equ} \label {eq:col-rules}
 \omega '=v_t~,\quad v_t'=\omega ~,\quad v_n'=-v_n~.
\end{equ}
This model was introduced in \cite{mejia,larralde} (with a different geometry)
to study macroscopic transport in terms of its microscopic dynamics.  It was
found that the disc mediated an effective interaction among the particles that
leads to a realistic macroscopic transport behavior.  Moreover, at each cell
this effective interaction is sufficient for the establishment of approximate
local thermal equilibrium \cite{mejia,eckmann-young}.

As a result of the interaction, when a particle enters the billiard from one
of the reservoirs, its probability to reach the other reservoir, \emph{i.e.},
the transmission coefficient, depends not only on the geometry of the billiard
but also on the local thermodynamic gradients. This dynamical effect was
studied in detail in \cite{eckmann-mejia} in a billiard of (effectively)
interacting particles with a geometry similar to the one that we describe
below (see also \cite{eckmann-young}).

\begin{figure}[!t]
\begin{center}
  \includegraphics[scale=0.4]{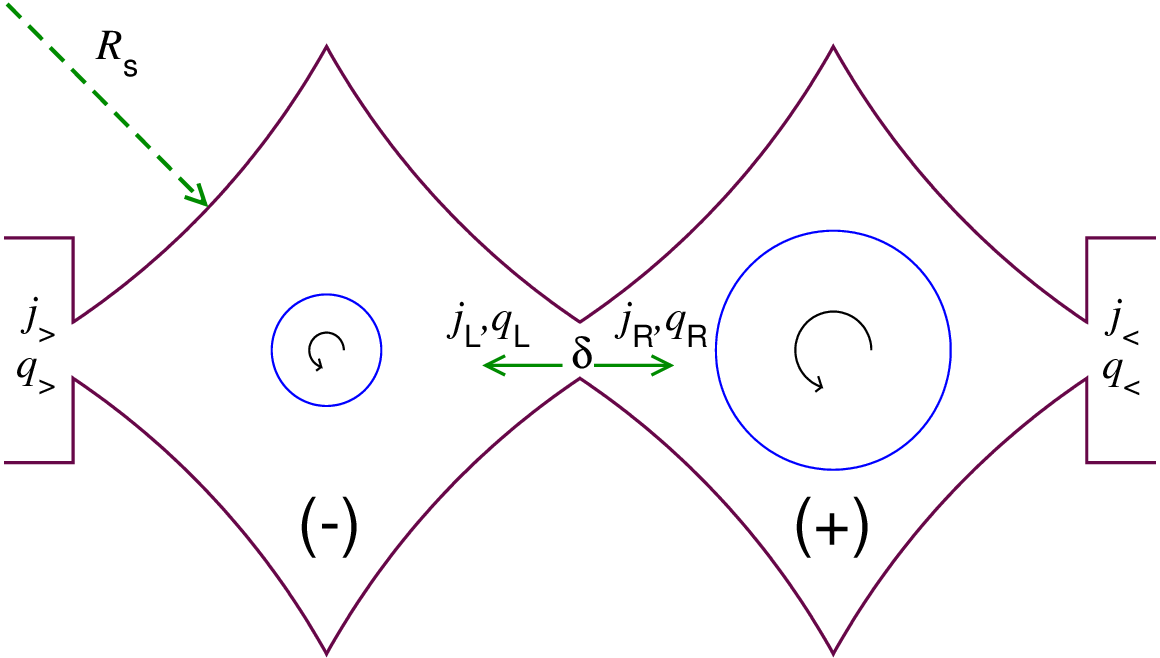}
\caption{
  Geometry of the rectifier. The boundary of each cell is made of four arcs of
  circles of  radius $R_s$. The radius of  the rotating discs is  chosen to be
  larger than  the width  of the  openings $\delta$, so  that no  particle can
  cross the cell  without undergoing any collision.  The  arrows represent the
  flows of particles across the openings  and the symbols fix the notation for
  the rates of  particles $j$ and of energy  $q$.  Unless otherwise specified,
  the numerical results in this letter correspond to a rectifier with discs of
  radius $0.04$  for the (-) cell  and $0.18$ for the  (+) cell, $\delta=0.04$
  and $R_s=1$.
 \label{fig:geometry}}
\end{center}
\end{figure}

To set up the thermal rectifier, we exploit the dependence of the dynamical
effect described above on the geometry of the cell.  We choose the size of the
disc in the left cell (-) smaller than the disc in the right cell (+).  The
transmission coefficient will then depend differently on the thermodynamic
fields, and this leads to TR.  The geometry of our setup is described in
Fig.~\ref{fig:geometry}.

To  impose nonequilibrium,  the two  cells are  connected at  the ends  to two
reservoirs  of particles  through openings  of  the same  size $\delta$.   The
reservoirs are  idealized as  infinite chambers containing  an ideal gas  at a
certain density $n$ and temperature $T$ \cite{larralde}.

From the left reservoir, particles are injected into the (-) cell at a
constant rate $j_>$ with Boltzmann distributed energy with a mean temperature
$T_>$.  Analogously, the right reservoir injects particles into the (+) cell
at rate $\j_<$ and temperature $T_<$. The rate at which energy is injected
into the system from the left and right reservoirs is then given by
$q_>=\frac{3}{2}j_>T_>$ and $q_<=\frac{3}{2}j_<T_<$ respectively
\cite{eckmann-young}. Each injected particle is eventually re-absorbed by one
of the reservoirs and this happens when it hits one of the reservoirs.

The density $n$ and temperature $T$  of the reservoir are related with its
injection rates as
\begin{equ} 
  n \propto \frac{j}{T^{1/2}} \propto \frac{j^{3/2}}{q^{1/2}} \ .
\end{equ}

At the junction between the two cells particles cross from (-) to (+) at a
rate $j_\R$ and from (+) to (-) at a rate $j_\L$. We denote by $q_\R$ and
$q_\L$ the corresponding energy rates.

In the steady state the conservation of particle number and of energy can be
expressed as a set of balance equations, \cite{eckmann-mejia}
\begin{equa}[2]\label{eq:balance}
j_\L \ &=& \ \frac{\aLR(1-\aLL)j_> + (1-\aRR)j_<}{1-\aLR\aRL} \ ,\\
q_\L \ &=& \ \frac{\bLR(1-\bLL)q_> + (1-\bRR)q_<}{1-\bLR\bRL} \ ,\\
\end{equa}
where $\aLL$ is the reflection probability for the particles that are injected
into the cell (-) from the left and $\aRL$ for the particles injected into the
cell (-) from  the right and similarly for the (+)  cell.  The parameters $\b$
stand for  the energy  reflection coefficients with  the same indexing  as the
$\a$'s. Analogous equations are obtained for $j_\R$ and $q_\R$, by
exchanging $\L\leftrightarrow\R$, $+\leftrightarrow -$, and
$>\,\leftrightarrow\, <$.

 Note  that out  of equilibrium  the reflection from  the left  is not
necessarily the  same to the  reflection from the  right due to  the effective
particle interaction.

In terms  of the  rates at the  junction the  particle current is  obtained by
$\Phi_n = j_\R -  j_\L$ and analogously, the heat current is  $\Phi_u = q_\R -
q_\L$. Using \eref{eq:balance} the currents are
\begin{equ} \label{eq:currents}
\Phi_n   =  \frac{\tau^\a_\L j_> - \tau^\a_\R j_<}{1 -  \aLR\aRL} \quad ; \quad
\Phi_u   =  \frac{\tau^\b_\L q_> - \tau^\b_\R q_<}{1 -  \bLR\bRL} \ ,
\end{equ}
where $\tau^x_y=\left(1-x^-_y\right)\left(1-x^+_y\right)$, with $x=\{\a,\b\} ,
y=\{\L,\R\}$, are the total transmission probabilities for the particles that
enter the channel from the $y$ side.  The currents in \eref{eq:currents} are
obtained as the difference of the products of these transmission probabilities
with the injection rates, weighted by  combinatorial factors that account for
multiple reflections between (-) and (+).

\noindent{\bf Reflection coefficients for a single cell}. In   \cite{eckmann-mejia}  a  phenomenological   theory  for   the  reflection
probabilities  $\alpha ^{\pm}_{\LR}$ was  obtained. These results
are for a single cell with
injection of  particles and energy at  rates $j_>$, $q_>$ from  the left and
$j_<$, $q_<$ from the right.  The reflection probabilities depend on the local
thermodynamic fields, that is
\begin{equ} \label{eq:alpha}
\a_\LR \equiv \a_\LR\left(j_>,q_>,j_<,q_<\right) ; \b_\LR \equiv
\b_\LR\left(j_>,q_>,j_<,q_<\right) \ .
\end{equ}
Since the dynamics is homogeneous in energy, the number of independent
variables in \eref{eq:alpha} is reduced to three.  It was found convenient to
distinguish two contributions: one (called $\aG$ and $\bG$), of purely
geometrical origin and another (called $\e$ and $\v$), of dynamical origin
arising as the system is driven out of equilibrium:
\begin{equa}[2] 
\a_\LR &=& \aG\left(n\right) +
\e_\LR\left(n,\delta j,\delta q\right) \ , \\
\b_\LR &=& \bG\left(n\right) +
\v_\LR\left(n,\delta j,\delta q\right) \ ,
\end{equa}
where $\delta j=(j_>-j_<)/(j_>+j_<)$ and similarly for $\delta q$.
By  definition,  $\varepsilon$  must vanish  at
equilibrium, {\em i.e.},  $\e_\LR(n,0,0) \ = \ 0$,  and by left-right symmetry
of the single cell, $\e_\L = -\e_\R$.

Detailed  analysis showed that:
\begin{itemize}
\item{}
At equilibrium  $\alpha_\LR = \aG(n)$.  
\item{}$\aG(n)$ saturates,  for low
and  for  high  densities,  to  two  constant values  connected  by  a  linear
dependence  for   intermediate  densities.  
\item{}  $\e$   vanishes  on  a
submanifold  of  $(j_>,q_>,j_<,q_<)$   where  $T_>=T_<$. This means
that {\em deviations from the equilibrium $\alpha $'s only occur for
  $\delta j\ne0$.}
\item{} Close to equilibrium, the dynamic component depends linearly on the
  local gradients of the particle injection, namely $\e_{\LR}=\pm\AA\delta j$,
  $\v_{\LR}=\pm\Bm\delta j$ Note that $\v$ depends on $\delta j$, \emph{not}
  on $\delta q$.
\end{itemize}

\begin{figure}[!t]
\begin{center}
\includegraphics[scale=0.45]{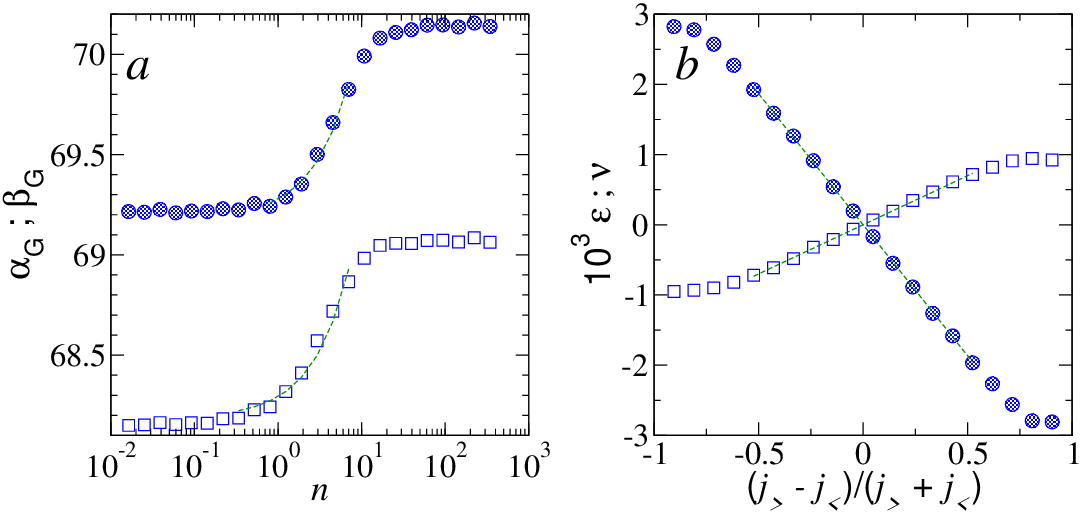}
\caption{
  $a$) Geometrical  reflections $\aG(n)$ and  $\bG(n)$ for a single  cell with
  the same dimensions as cell  (+), maintained at equilibrium.  $b$) Dynamical
  reflections $\e_\L$  and $\nu_\L$  as a function  of the local  gradients of
  particle injection rates  for the cell (+).  Note that  by definition $\aG =
  (\alpha_\L  + \alpha_\R)/2$  and similar  for  $\bG$.  In  both panels,  the
  symbols  correspond to  the energy  reflection (hashed  circles) and  to the
  particle reflection (open squares) and the dashed lines correspond to linear
  fits.
 \label{fig:phenom}}
\end{center}
\end{figure}

In Fig.~\ref{fig:phenom}, we show the behavior of these two contributions for
a cell of type (+). These results were obtained under very general reasoning
and thus, they are expected to hold qualitatively for different models of
interactions as long as the dynamics inside each cell is sufficiently mixed
and phase-space volume is preserved \cite{eckmann-mejia}.

We have  verified that  this qualitative behavior  is the same  for different
sizes  of the  disc,  the only  change  being the  values  for the  saturation
plateaus of the geometric contribution (Fig.~\ref{fig:phenom}-a) and the slope
of the linear dependence of  the dynamical contribution on the local gradients
(Fig.~\ref{fig:phenom}-b). In  particular we have  found (not shown)  that the
variation,  from the  lower to  the upper  plateau, of  $\aG$ is  much smaller
($\sim20$ times) for the (-) cell than for the (+) cell and similar for $\bG$.
Therefore, these quantitative values characterize the geometry of the cell and
its dynamical response to the presence of thermodynamic gradients.

\noindent{\bf Combining the two cells}. We now apply these formulas
for the two-cell case: For the (-) cell, the particle currents are
$j_>$ and $j_\L$, while for the (+), they are $j_\R$ and $j_<$. From
our findings for a single cell, we find, to a very good approximation:
\begin{equ} \label{eq:j-phenom}
\eLm = -\eRm = \Am \frac{j_> - j_\L}{j_> + j_\L} ; \quad \eLp =-\eRp= \Ap \frac{j_R -
  j_<}{j_R + j_<} \ ,
\end{equ}
\begin{equ} \label{eq:q-phenom}
\vLm =-\vRm= \Bm \frac{j_> - j_\L}{j_> + j_\L} ; \quad \vLp = -\vRp=\Bp \frac{j_R -
  j_<}{j_R + j_<} \ .
\end{equ}
Here and below $\pm$ refers to one of the two cells.  Substituting
\eref{eq:j-phenom} into \eref{eq:balance} we obtain the currents $j_\LR$ at
the center of the cell in terms of the parameters that define the geometry of
the cells as
\begin{equa}[2] \label{eq:j-sol}
j_\L \ &=& \ \frac{\rap(1-\ram)j_> + (1-\rap)j_<}{1-\ram\rap} \ ,\\
j_\R \ &=& \ \frac{(1-\ram)j_> + \ram(1-\rap)j_<}{1-\ram\rap} \ ,
\end{equa}
with effective   reflection   coefficients  given   by
$\varrho_\alpha^\pm  =  \aG(n^\pm) -  \mathcal{A}^\pm$  and with $n^\pm$ the  mean
density in the respective cell. One finds finally for the particle current
(to within our approximation)
\begin{equ} \label{eq:j-phi+}
\Phi_n = -\frac{\left(1-\ram\right)\left(1-\rap\right)}{1 - \ram\rap}\left(j_<
  - j_>\right) \ .
\end{equ}
Thus, near equilibrium, $\Phi_n$ does \emph{not}  depend on the energy
field $q$. In  contrast, from \eref{eq:q-phenom} it is  clear that the current
of  \emph{energy}  depends  on  both  fields.    Substituting  \eref{eq:j-sol}  into
\eref{eq:q-phenom} we find for the dynamical reflections
\begin{equ} \label{eq:nu-PG}
\v_\L = \frac{\Bm}{1 + \frac{2\left(1-\ram\rap\right)j_>}{\left(1-\rap\right)\left(j_<-j_>\right)}} ~, \quad
\v_\R = \frac{\Bp}{-1 + \frac{2\left(1-\ram\rap\right))j_<}{\left(1-\ram\right)\left(j_<-j_>\right)}} \ .
\end{equ}
Finally,     inserting    \eref{eq:q-phenom}     and     \eref{eq:nu-PG}    in
\eref{eq:currents},  one   obtains  an  expression  for   the  energy  current
exclusively         in         terms         of         the         parameters
$\Phi_u=\Phi_u(j_>,q_>,j_<,q_<;\r_\a^\pm,\bG^\pm,\B^\pm)$.

\noindent{\bf Inverting the imposed fields}.  From now on we refer to
the case in which the injections at  the left are $(j_>,q_>)$ and at the right
$(j_<,q_<)$ as  positive gradient  (PG) and to  the inverted case  as negative
gradient (NG).   Due to  the effective interaction  among particles  the local
fields in the cells  for the NG are different from those of PG. 
Intuitively, this is because the temperature at the junction of the
two cells changes when one goes from PG to NG.
In particular,
the energy reflection coefficients  in \eref{eq:nu-PG} are not invariant
with respect to the sign of the  gradient.  
Let $n^{\pm}$ be the particle density for PG and let $\tilde n^{\pm}$
be the particle densities for the NG. Defining $\tilde{\r}_\a^\pm =
\a_\G(\tilde{n}^\pm) - \A^\pm$ and using the tilde quantities in
\eref{eq:nu-PG}, one obtains in analogy with \eref{eq:currents} for the NG:
\begin{equ}
\tilde{\Phi}_u   =  \frac{\tilde{\tau}^\b_\L q_< - \tilde{\tau}^\b_\R q_>}{1 -  \tilde{\b}_\L^+\tilde{\b}_\R^-} ~.
\end{equ}
with
$\tilde{\tau}^\b_{\LR}=(1-\tilde{\b}_{\LR}^-)(1-\tilde{\b}_{\L,R}^+)$
and $\tilde{\b}_{\LR}^\pm = \tilde{\b}_\G^\pm \pm \tilde{\v}^\pm$.

%  and where
%$\tilde{\tau}^\b_{\LR}=(1-\tilde{\b}_{\LR}^-)(1-\tilde{\b}_{\L,R}^+$),
%$\tilde{\b}_{\LR}^\pm = \tilde{\b}_\G^\pm \pm \tilde{\v}^\pm$,
%\begin{equ} 
%\tilde{\v}^- = \frac{\B^-}{1 + \frac{2\left(1-\tilde{\r}_\a^-\tilde{\r}_\a^+\right)j_<}{\left(1-\tilde{\r}_\a^+\right)\left(j_<-j_>\right)}} \quad , \quad
%\tilde{\v}^+ = \frac{\B^+}{-1 +
%  \frac{2\left(1-\tilde{\r}_\a^-\tilde{\r}_\a^+\right))j_>}{\left(1-\tilde{\r}_\a^-\right)\left(j_<-j_>\right)}} \ .
%\end{equ}
%and $\tilde{\r}_\a^\pm = \a_\G(\tilde{n}^\pm) - \A^\pm$.
%
To quantify the TR we define the thermal rectification index as the quotient
between the heat currents for the PG and NG
\begin{equ}
\Theta_u = \frac{\mathrm{max}\{|\Phi_u|,|\tilde{\Phi}_u|\}}{\mathrm{
    min}\{|\Phi_u|,|\tilde{\Phi}_u|\}}  \ ,
\end{equ}
so that $\Theta_u\ge1$. For zero rectification $\Theta_u=1$.

\begin{figure}
\begin{center}
\includegraphics[scale=0.45]{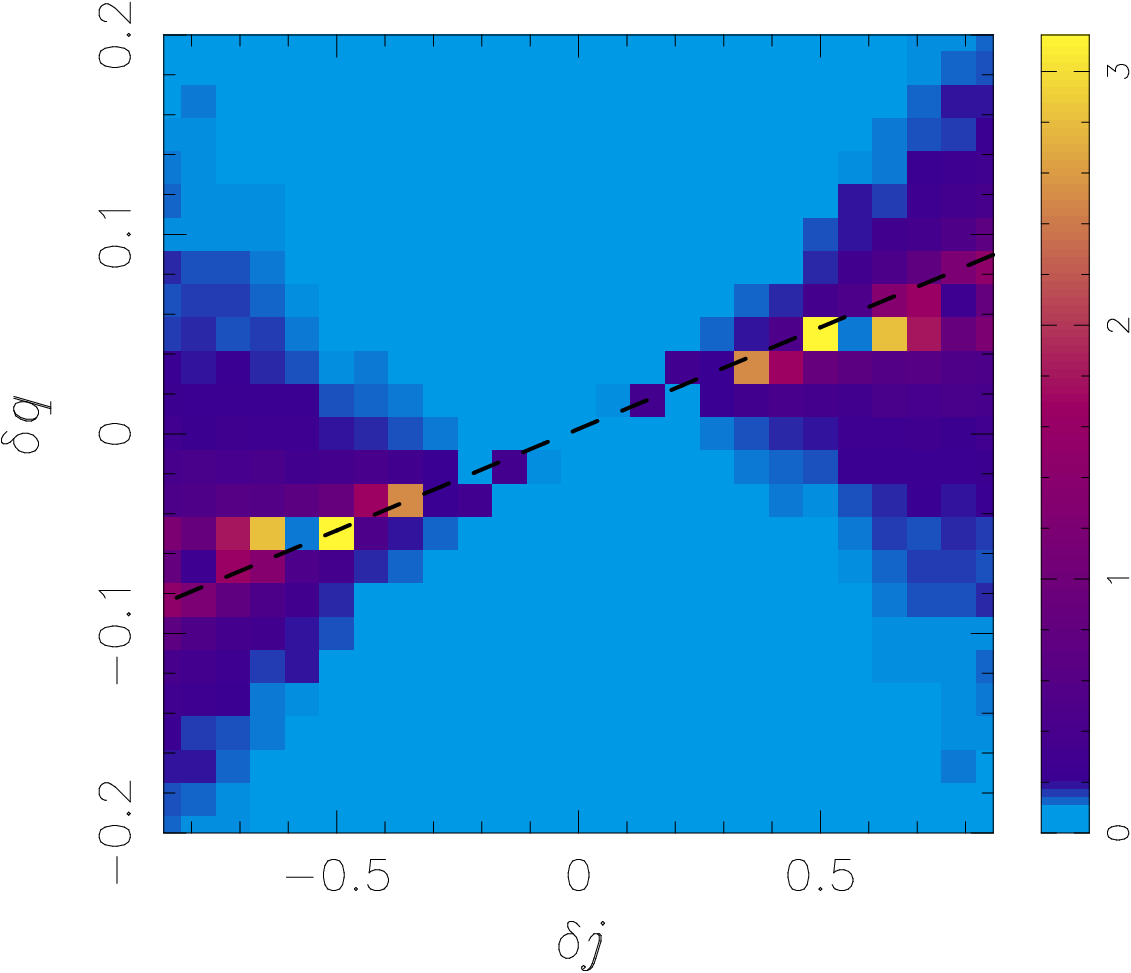}
\caption{Color density plot of the logarithm of the thermal rectification index
  $\log\left(\Theta_u\right)$ as  a function of the  external gradients $\delta
  j$ and $\delta q$ for  the  billiard
  described in  Fig.~\ref{fig:geometry}. By construction the  data is symmetric
  along the diagonal. The dashed line  correspond fit to the numerical data of
  the zero-current condition \eref{eq:zero-current}.
 \label{fig:thermal}}
\end{center}
\end{figure}

\noindent{\bf Results}. In  Fig.~\ref{fig:thermal}, we show  in a  density plot  the logarithm  of the
thermal rectification index $\Theta_u$ as a function of the external gradient.
A large TR  is found in a  region where $\delta  j$ and $\delta q$  are linearly
related. The  maximal value  of $\Theta_u$ observed  in Fig.~\ref{fig:thermal}
corresponds to heat currents $\Phi_u$  and $\tilde{\Phi}_u$ where one is three
orders of magnitude smaller than the other.

As  found in  previous models,  the magnitude  of TR  depends on  the external
gradients.   For our  type  of models  the  largest values  of $\Theta_u$  are
obtained when either $\Phi_u$ or $\tilde{\Phi}_u$ (but no both) is very small.
This means  that TR is maximal when  for a given external  gradient the system
behaves  as an  insulator.  If  for a  given gradient  $(\delta  j,\delta q)$,
$\Phi_u=0$   then   for  the   inverted   gradient   $(-\delta
j,-\delta   q)$, 
$\tilde{\Phi}_u$ will not in general  be zero. This is because the reflection
probabilities  will  be  different  through  their  dependence  on  the  local
thermodynamic fields.

Equating $\Phi_u$ to  zero we obtain a condition  that involves the reflection
probabilities $\b$:
\begin{equ} \label{eq:zero-current}
\frac{(1-\bLL)(1-\bLR)}{(1-\bRL)(1-\bRR)} = \frac{q_<}{q_>} \ .
\end{equ}
This equation states  the values of the external gradient  for which our model
behaves  as  a heat  insulator.   The  dashed  line in  Fig.~\ref{fig:thermal}
corresponds to  a linear fit to the  condition \eref{eq:zero-current} computed
from the values obtained from the numerical simulation.

\noindent{\bf Discussion and conclusions}. In our  model TR  is due to  the dynamical  response of each  cell to  a given
external gradient and thus, to the interacting character of the particles. If
both cells and their disks are identical, then the response for the PG is
identical to the one for the NG and there is no TR. On the other hand, it is
tempting to consider the opposite case when the domains are equal, but the
cell (-) has no disc at all.  While we have observed TR for this case, we do
not report further on this as our analytical treatment is not longer valid.

As we have seen before, rectification has a geometric component
(occurring when $\delta j=0$).
For   the  billiard  of   Fig.~\ref{fig:geometry}  a  \emph{thermal}
rectification  of   $\Delta_u\approx1.04$  is  obtained   when  the  geometric
component for the  PG and the NG  takes values in the lower  and upper plateaus
respectively. 

This rectification power can be enhanced to much larger values
when the billiard behaves as an insulator for positive or negative
gradient, and this can be achieved when $\delta j\ne0$.

The geometric component also allows a rectification of  the
 \emph{particle}
current. 
However, this rectification cannot be enhanced as from \eref{eq:j-phi+}
it is clear that the particle  current $\Phi_n$ is zero only when $\delta j=0$.

For  open billiards like  the one  studied here,  the finite  particle current
contributes  to the  TR only  through  the dynamical  component $\v^\pm$.   In
practice,  this situation  corresponds to  the transport  in  porous materials
where  the energy is  carried by  the particles.   However, for  the analogous
closed billiard for  which only energy is exchange between  the system and the
reservoirs large thermal rectification have been numerically observed.

In  this  Letter, we  have  established  the  possibility to  observe  thermal
rectification in  billiard-like mechanical systems.   We have shown  that when
the gas of particles inside the  billiard interact, the response of the system
to  an external  thermodynamic gradient  is dynamically  controlled.   We have
studied this dynamical  effect in terms of the  particle and energy reflection
coefficients for which an  analytical phenomenological formulation in terms of
the local  thermodynamic gradients is possible. This  dynamical response leads
to possibility that the system behaves as an insulator for one gradient and as
conductor   with  the  inverted   gradient  in   which  case,   large  thermal
rectification is  obtained.  Our results lead  to conjecture that  TR would be
observable in nanoscopic billiard-like systems  for which the gas of electrons
is not free.

\begin{acknowledgments}
  The  authors  acknowledge fruitful  discussions  with  Toma\u  z Prosen  and
  support by the Fonds National  Suisse.  CMM thanks the hospitality of Centro
  Internacional de Ciencias in Cuernavaca  Mexico, where part of this work was
  done.
\end{acknowledgments}

\end{document}